\documentclass{PoS}

\title{Electron contribution to $(g-2)_\mu$ at four loops}

\ShortTitle{Electron contribution to $(g-2)_\mu$ at four loops}

\author{Alexander~Kurz\\
        Institut f{\"u}r Theoretische Teilchenphysik, Karlsruhe Institute of
        Technology (KIT),\\ 76128 Karlsruhe, Germany\\
        E-mail: \email{alexander.kurz2@kit.edu}}

\author{Tao~Liu\\
        Department of Physics, University of Alberta, Edmonton AB T6G 2J1, Canada\\
        E-mail: \email{ltao@ualberta.ca}}

\author{Peter~Marquard\\
        Deutsches Elektronen Synchrotron DESY, Platanenallee 6, 15738 Zeuthen, Germany\\
        E-mail: \email{peter.marquard@desy.de}}

\author{Alexander V.~Smirnov\\
        Scientific Research Computing Center, Moscow State University, 119991,
        Moscow, Russia\\
        E-mail: \email{asmirnov80@gmail.com}}

\author{Vladimir A.~Smirnov\\
        Skobeltsyn Institute of Nuclear Physics of Moscow State University, 119991,
        Moscow, Russia\\
        E-mail: \email{smirnov@theory.sinp.msu.ru}}

\author{\speaker{Matthias Steinhauser}\\
        Institut f{\"u}r Theoretische Teilchenphysik, Karlsruhe Institute of
        Technology (KIT),\\ 76128 Karlsruhe, Germany\\
        E-mail: \email{matthias.steinhauser@kit.edu}}

\abstract{In this contribution we summarize the recent calculation of the 
  complete electron contribution to the anomalous magnetic moment of the muon
  at four-loop order.}

\FullConference{Loops and Legs in Quantum Field Theory\\
		24-29 April 2016\\
		Leipzig, Germany}

\graphicspath{ {figs/} }
\usepackage{amstext}

\begin{document}

\section{Introduction}

The anomalous magnetic moment of the muon is among the most precisely known
quantities in particle physics. At the same time there is a long-standing
deviation between the experimental measurement and the theory prediction which
amounts to about three standard deviations.  On the experimental side there
are upcoming new experiments which either use the same method as in the E821
experiment at BNL~\cite{Bennett:2006fi,Roberts:2010cj} but reduce the
uncertainties by about a factor four, or even use a completely different
technique which would eliminate doubts on possible systematic effects (for
details see, e.g., Ref.~\cite{Hertzog:2015jru}).

On the theory side it is certainly necessary to improve on the hadronic
contributions, both from the vacuum polarization and from light-by-light-type
diagrams.  Furthermore, it is mandatory to cross check the four-loop QED
contribution since the full result has only been obtained by one
group~\cite{Kinoshita:2004wi,Aoyama:2007mn,Aoyama:2012wk}. In a series of
works~\cite{Kurz:2013exa,Kurz:2015bia,Kurz:2016bau} the fermionic pieces 
have been confirmed. The cross check of the purely photonic
part is still missing. In this contribution we report on the calculation and
results of the diagrams involving closed electron loops~\cite{Kurz:2016bau}.

\section{The method}

\begin{figure}[b]
  \begin{center}
    \begin{tabular}{cccc}
      \includegraphics[scale=0.30]{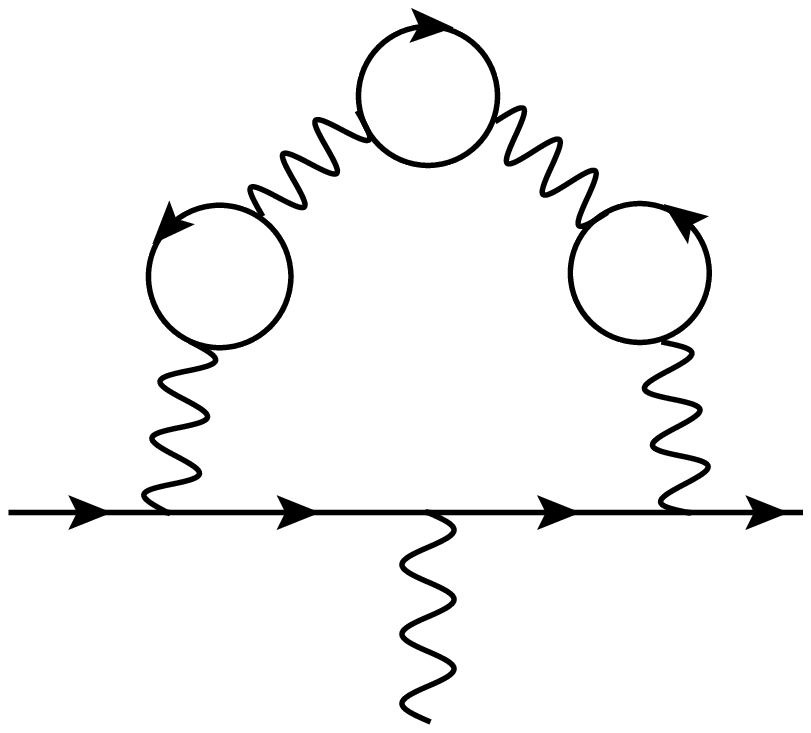}
      & \includegraphics[scale=0.30]{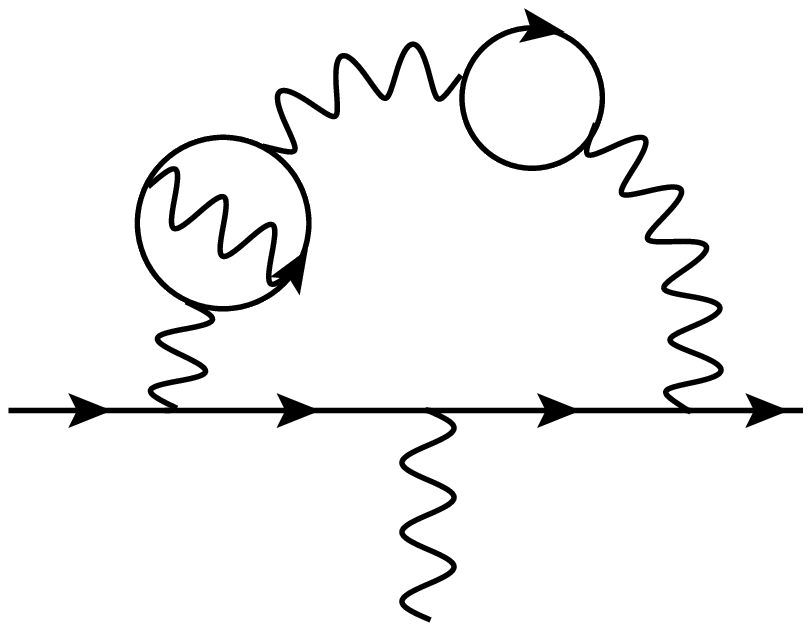}
      & \includegraphics[scale=0.30]{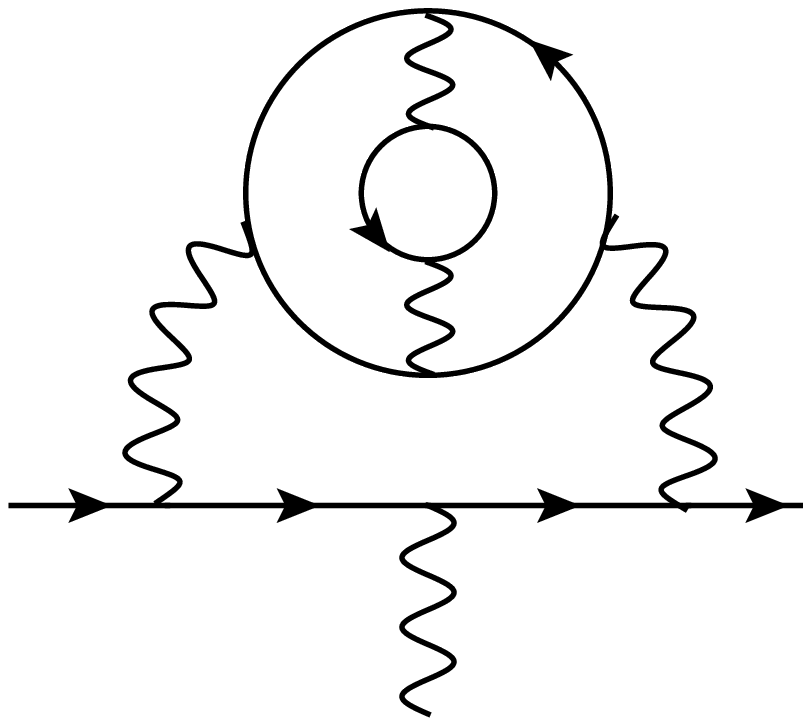}
      & \includegraphics[scale=0.30]{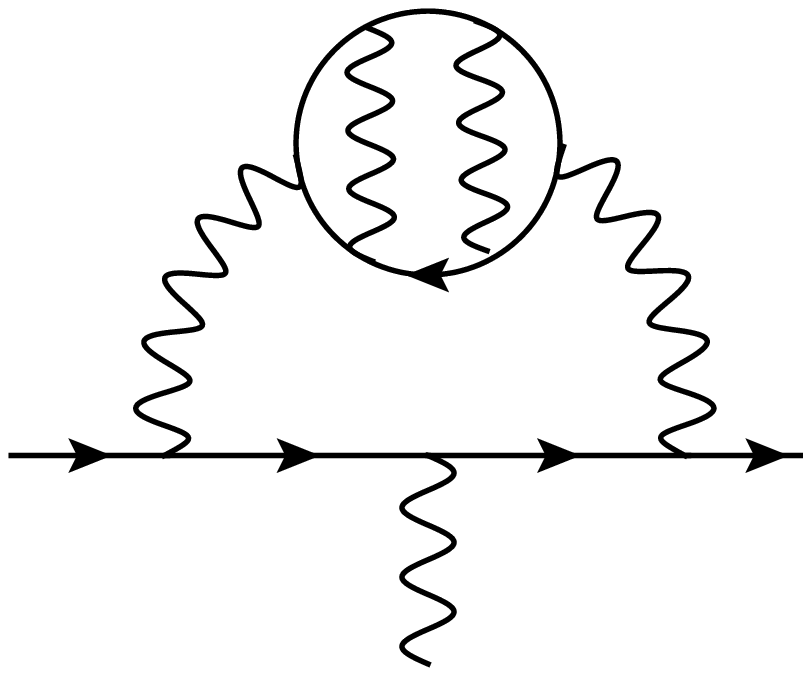} \\
      I(a) & I(b) & I(c) & I(d) \\
      \includegraphics[scale=0.30]{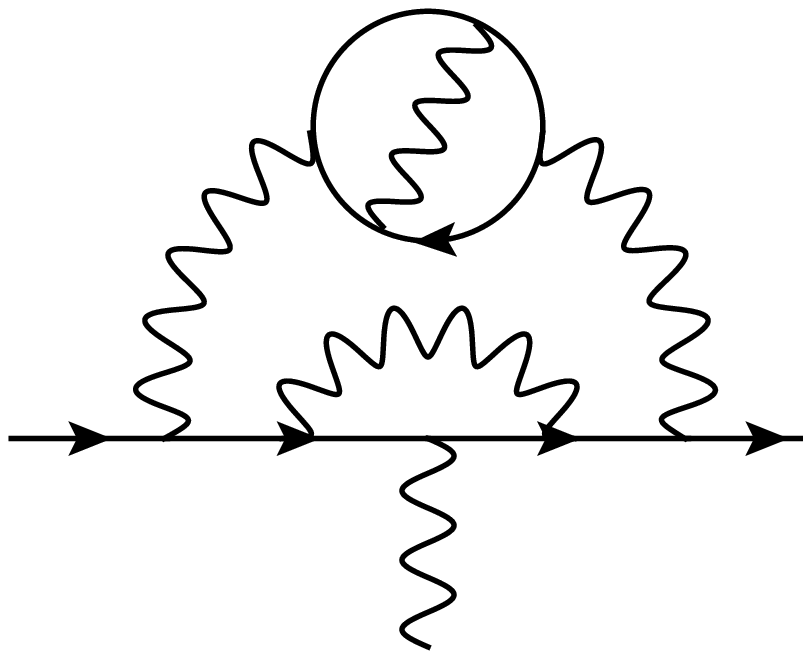}
      & \includegraphics[scale=0.30]{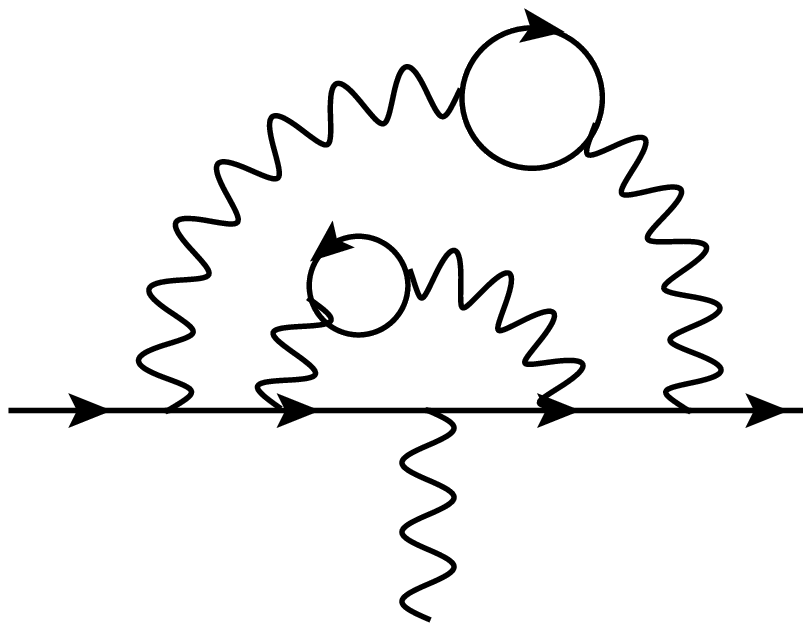}
      & \includegraphics[scale=0.30]{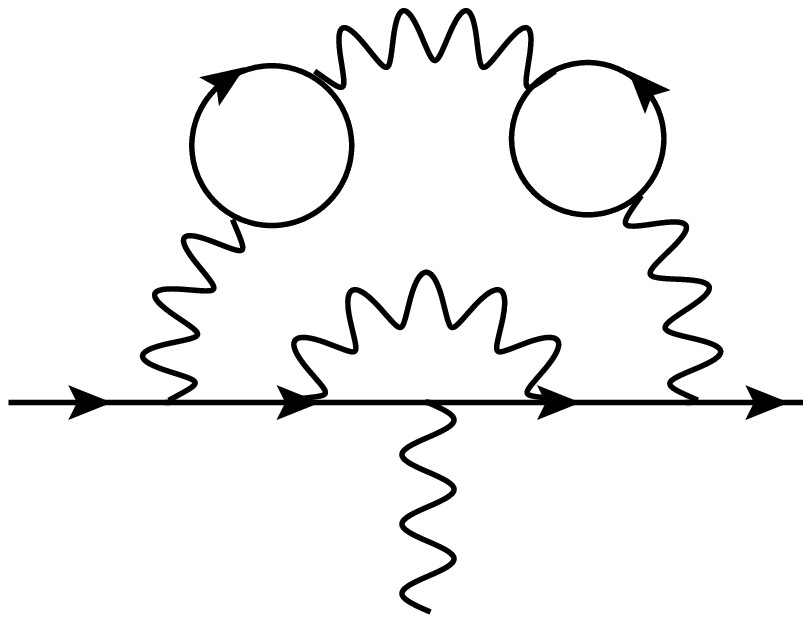}
      & \includegraphics[scale=0.30]{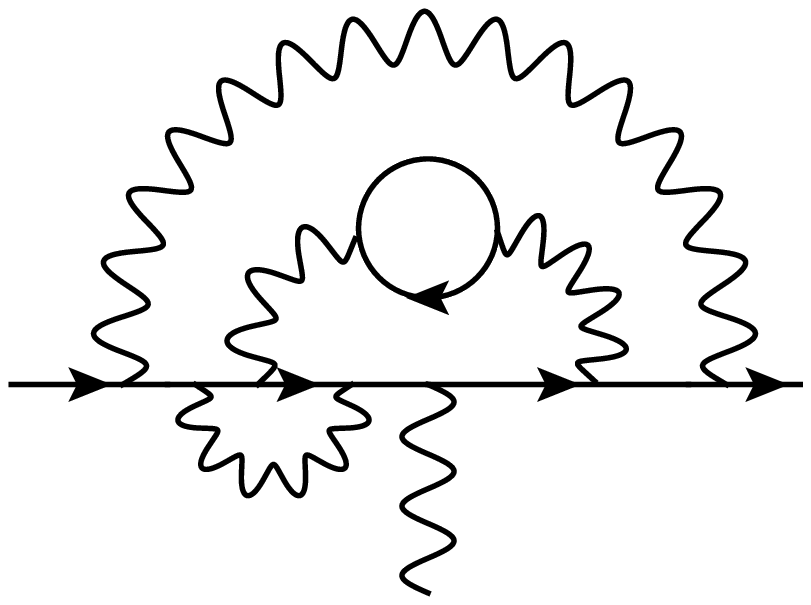} \\
      II(a) & II(b) & II(c) & III \\
      \includegraphics[scale=0.30]{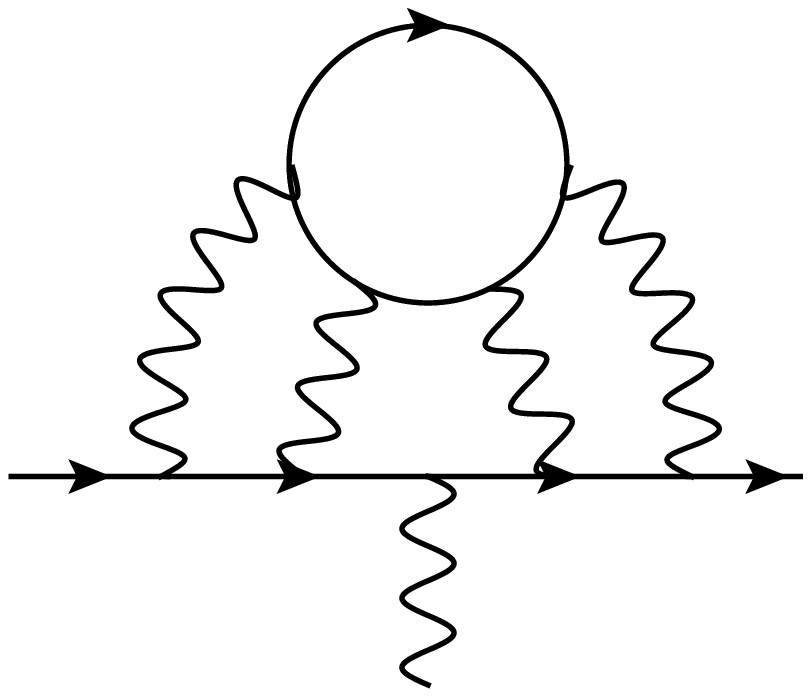}
      & \includegraphics[scale=0.30]{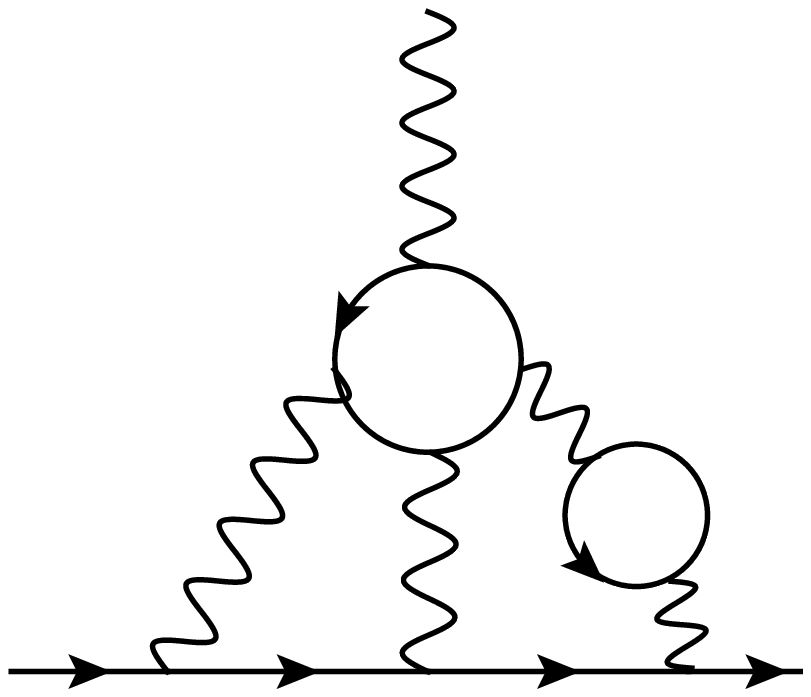}
      & \includegraphics[scale=0.30]{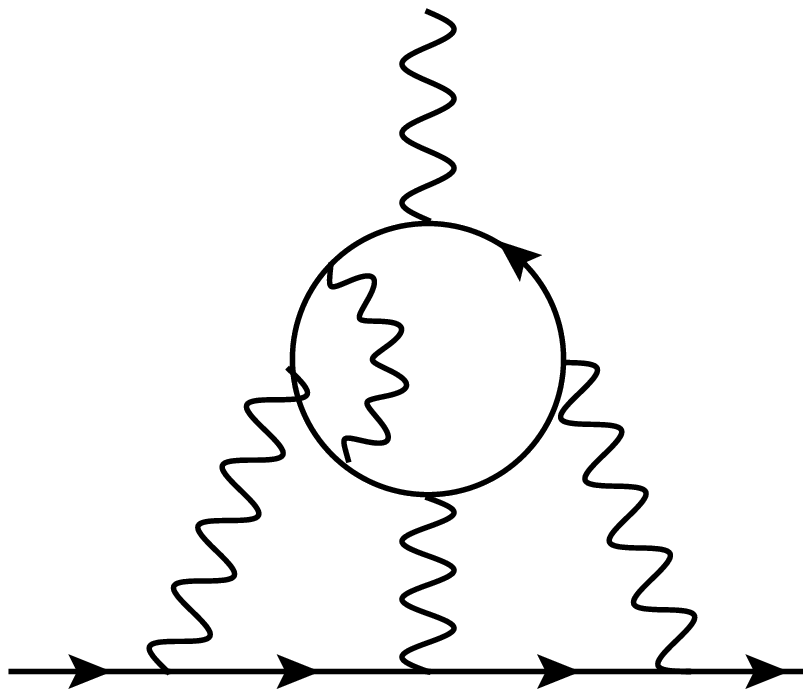}
      & \includegraphics[scale=0.30]{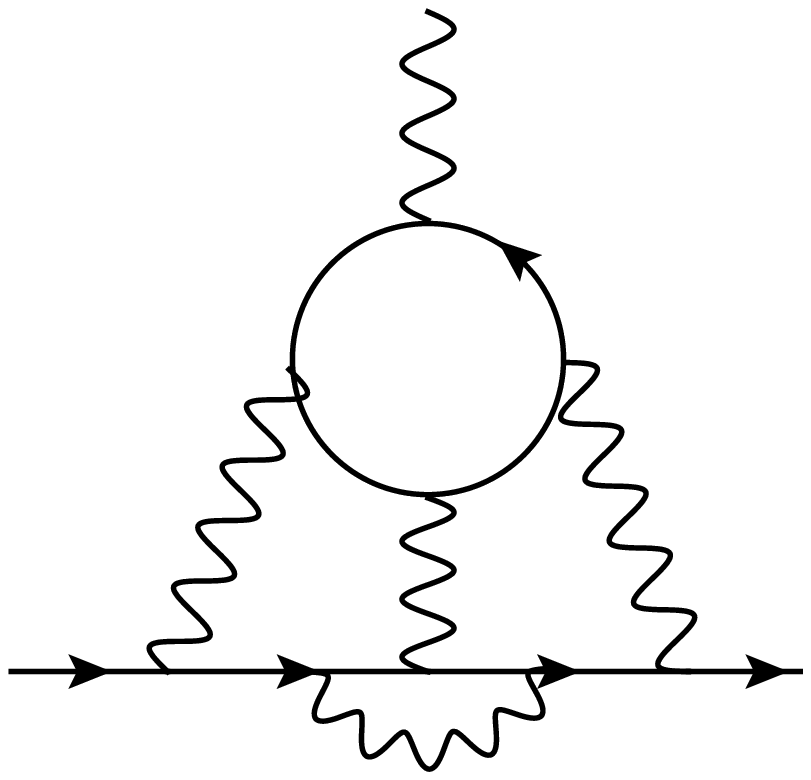}  \\
      IV(d) & IV(a) & IV(b) & IV(c)
    \end{tabular}
  \end{center}
  \caption{
    Four-loop diagram classes for $a_\mu$ 
    containing at least one closed electron loop. The
    external solid lines represent muons, the solid loops denote
    electrons, muons or taus, and the wavy lines represent photons.}
  \label{fig::classes}
\end{figure}

It is convenient to sub-divide the contributing four-loop diagrams in twelve
classes~\cite{Aoyama:2012wk} which are introduced in Fig.~\ref{fig::classes}
with the help of sample Feyman diagrams.

The approach used for the computation of the four-loop diagrams differs from
the one applied in Ref.~\cite{Aoyama:2012wk} in many ways.
In~\cite{Kurz:2016bau} we generate in a first step amplitudes for each
individual vertex diagram which contain both the electron ($m_e$) and the muon
mass ($m_\mu$). At this point we exploit the fact that $m_e\ll m_\mu$ and
apply an asymptotic expansion which expresses each amplitude into a sum of
so-called sub-diagrams. Each sub-diagram is written as a product of one-scale
integrals which are much easier to compute.

The various sub-diagrams involve different types of integrals which have to be
treated separately. Most of them are well studied in the literature and
analytic results can be obtained. However, there are two types where this is not
the case: four-loop on-shell integrals and four-loop integrals involving
propagators of the form $1/(2 \ell \cdot q)$ where $q$ is the external
momentum with $q^2=m_\mu^2$, in the following called ``linear integrals''.  

At this point we apply an appropriate projector to the magnetic form factor
and expand afterwards in the photon momentum to obtain the static limit.  Then
we perform the traces and decompose each amplitude into a sum of scalar
integrals. For simple integral types (like two-loop vacuum integrals) we
can directly insert the analytic results for the integrals. The more complicated
ones are reduced to so-called master integrals using the program packages {\tt
  FIRE}~\cite{Smirnov:2014hma} and {\tt crusher}~\cite{crusher}.
In this way we obtain an analytic result for the muon anomalous magnetic moment
in term of a relatively small number [${\cal O}(100)$] of master integrals.
This is the case for all coefficients of $(m_e/m_\mu)^n$ (we expanded up to $n=3$).
Note that as far as the four-loop master integrals are concerned
the odd powers of $m_e/m_\mu$ only involve linear integrals while
the even powers get contributions from on-shell and linear integrals.

It is only at this point when we pass on to numerical methods since
to date not all master integrals are available in analytic form.
This is the origin of the uncertainty in our final results, see below.

The numerical evaluation of the four-loop on-shell master
integrals is described in detail in Ref.\cite{Marquard:2016dcn}. 
A similar approach has also been used for the linear integrals, see
also Ref.~\cite{Kurz:2016bau}.

\section{Results for $(g-2)_\mu$}

In this section we present results for the anomalous magnetic moment
of the muon. We cast the perturbative expansion in the form
\begin{eqnarray}
  \frac{(g-2)_\mu}{2} \,\,=\,\, a_\mu &=& 
  \sum_{n=1}^\infty a_\mu^{(2n)} \left( \frac{\alpha}{\pi} \right)^n
  \,,
  \label{eq::amu}
\end{eqnarray}
where $n$ counts the number of loops. $a_\mu^{(2n)}$ is conveniently split
into several pieces according to the particles present in the loop.
In particular, we have for the four-loop term
\begin{eqnarray}
  a_\mu^{(8)} &=& A_1^{(8)} + A_2^{(8)}(m_\mu / m_e) + A_2^{(8)}(m_\mu / m_\tau) 
  + A_3^{(8)}(m_\mu / m_e, m_\mu / m_\tau)
  \,,
  \label{eq::Amu}
\end{eqnarray}
where $A_1^{(8)}$ denotes the universal part which includes the pure photonic
corrections and closed muon loops.  $A_2^{(8)}(m_\mu / m_e)$ ($A_2^{(8)}(m_\mu
/ m_\tau)$) contains in addition at least one closed electron (tau) loop and
$A_3^{(8)}(m_\mu / m_e, m_\mu / m_\tau)$ contains at least one electron and
one tau loop.

In Table~\ref{tab::res} the results from the individual diagram classes
contributing to $A_2^{(8)}(m_\mu / m_e)$ are shown. For practical reasons 
only the sum is presented for the classes I(b)+I(c) and II(b)+II(c)
and a further splitting is carried out in case more than one electron loop is
present (see Ref.~\cite{Kurz:2016bau} for a detailed discussion.)

\begin{table}[tb]
\begin{center}
\begin{tabular}{l@{\hskip -0.2cm}lll}
$A_2^{(8)}(m_\mu/m_e)$ & \hphantom{-00} \cite{Kurz:2016bau,Kurz:2015bia} & \hphantom{-00} literature & \\
\hline
I(a0) & $\hphantom{-00}7.223076$ & $\hphantom{-00}7.223077 \pm 0.000029$ & \cite{Kinoshita:2004wi} \\
 & & $\hphantom{-00}7.223076$ & \cite{Laporta:1993ds} \\
I(a1) & $\hphantom{-00}0.494072$ & $\hphantom{-00}0.494075 \pm 0.000006$ & \cite{Kinoshita:2004wi} \\
 & & $\hphantom{-00}0.494072$ & \cite{Laporta:1993ds} \\
I(a2) & $\hphantom{-00}0.027988$ & $\hphantom{-00}0.027988 \pm 0.000001$ & \cite{Kinoshita:2004wi} \\
 & & $\hphantom{-00}0.027988$ & \cite{Laporta:1993ds} \\
I(a) & $\hphantom{-00}7.745136$ & $\hphantom{-00}7.74547 \pm 0.00042$ & \cite{Aoyama:2012wk} \\
\hline
I(bc0) & $\hphantom{-00}8.56876 \pm 0.00001$ & $\hphantom{-00}8.56874 \pm 0.00005$  & \cite{Kinoshita:2004wi} \\
I(bc1) & $\hphantom{-00}0.1411 \pm 0.0060$ & $\hphantom{-00}0.141184 \pm 0.000003$  & \cite{Kinoshita:2004wi} \\
I(bc2) & $\hphantom{-00}0.4956 \pm 0.0004$ & $\hphantom{-00}0.49565 \pm 0.00001$  & \cite{Kinoshita:2004wi} \\
I(bc) & $\hphantom{-00}9.2054 \pm 0.0060$ & $\hphantom{-00}9.20632 \pm 0.00071$ & \cite{Aoyama:2012wk} \\
\hline
I(d) & $\hphantom{0}\text{$-$}\hphantom{0}0.2303 \pm 0.0024$ & $\hphantom{0}\text{$-$}\hphantom{0}0.22982 \pm 0.00037$ & \cite{Aoyama:2012wk} \\
 & & $\hphantom{0}\text{$-$}\hphantom{0}0.230362 \pm 0.000005$ & \cite{Baikov:1995ui} \\
\hline
II(a) & $\hphantom{0}\text{$-$}\hphantom{0}2.77885$ & $\hphantom{0}\text{$-$}\hphantom{0}2.77888 \pm 0.00038$ & \cite{Aoyama:2012wk} \\
 & & $\hphantom{0}\text{$-$}\hphantom{0}2.77885$ & \cite{Laporta:1993ds} \\
\hline
II(bc0) & $\hphantom{0}\text{$-$}12.212631$ & $\hphantom{0}\text{$-$}12.21247 \pm 0.00045$ & \cite{Kinoshita:2004wi} \\
II(bc1) & $\hphantom{0}\text{$-$}\hphantom{0}1.683165 \pm 0.000013$ & $\hphantom{0}\text{$-$}\hphantom{0}1.68319 \pm 0.00014$ & \cite{Kinoshita:2004wi} \\
II(bc) & $\hphantom{0}\text{$-$}13.895796 \pm 0.000013$ & $\hphantom{0}\text{$-$}13.89457 \pm 0.00088$ & \cite{Aoyama:2012wk} \\
\hline
III & $\hphantom{-0}10.800 \pm 0.022$ & $\hphantom{-0}10.7934 \pm 0.0027$ & \cite{Aoyama:2012wk} \\
\hline
IV(a0) & $\hphantom{-}116.76 \pm 0.02$ & $\hphantom{-}116.759183  \pm 0.000292$ & \cite{Kinoshita:2004wi} \\
 & & $\hphantom{-}111.1 \pm 8.1$ & \cite{Calmet:1975tw} \\
 & & $\hphantom{-}117.4 \pm 0.5$ & \cite{Chlouber:1977dr} \\
IV(a1) & $\hphantom{-00}2.69 \pm 0.14$ & $\hphantom{-00}2.697443 \pm 0.000142$ & \cite{Kinoshita:2004wi} \\
IV(a2) & $\hphantom{-00}4.33 \pm 0.17$ & $\hphantom{-00}4.328885 \pm 0.000293$ & \cite{Kinoshita:2004wi} \\
IV(a) & $\hphantom{-}123.78\pm 0.22$ & $\hphantom{-}123.78551 \pm 0.00044$ & \cite{Aoyama:2012wk} \\
\hline
IV(b) & $\hphantom{0}\text{$-$}\hphantom{0}0.38 \pm 0.08$ & $\hphantom{0}\text{$-$}\hphantom{0}0.4170 \pm 0.0037$ & \cite{Aoyama:2012wk} \\
IV(c) & $\hphantom{-00}2.94 \pm 0.30$  & $\hphantom{-00}2.9072 \pm 0.0044$ & \cite{Aoyama:2012wk} \\
\hline
IV(d) & $\hphantom{0}\text{$-$}\hphantom{0}4.32 \pm 0.30$ & $\hphantom{0}\text{$-$}\hphantom{0}4.43243 \pm 0.00058$ & \cite{Aoyama:2012wk} \\
\hline
\end{tabular}
\end{center}
\caption{Final results for the different classes and comparison
  with the literature.}
\label{tab::res}
\end{table}

It is interesting to note that in some cases our coefficients have smaller
uncertainties (e.g. II(bc)) whereas for others we have obtained 
an uncertainty which is much worse than the one of~\cite{Aoyama:2012wk}
(e.g. IV(c) or IV(d)).
This can be traced back to complicated master integrals which at the
moment can only be evaluated with a few-digit precision. Let us stress
that, if necessary, the precision of our result can be improved systematically.

Our final result for $A_2^{(8)}(m_\mu/m_e)$ is given by
\begin{eqnarray}
  A_2^{(8)} &=& 126.34(38) + 6.53(30) = 132.86(48)\,,
\end{eqnarray}
where the first number after the first equality sign originates from the
light-by-light-type diagrams IV(a), IV(b) and IV(c).
Our final numerical uncertainty amounts to approximately ${0.5} \times
(\alpha/\pi)^4 \approx {1.5} \times 10^{-11}$. It is larger than
the uncertainty in Ref.~\cite{Aoyama:2012wk}.  Nevertheless it is
sufficiently accurate as can be seen by the comparison to the
difference between the experimental result and theory prediction which
is given by~\cite{Aoyama:2012wk}
\begin{eqnarray}
  a_\mu({\rm exp}) - a_\mu({\rm SM}) &\approx& 249(87) \times 10^{-11}
  \,.
  \label{eq::amu_diff}
\end{eqnarray}
Note that the uncertainty in Eq.~(\ref{eq::amu_diff}) receives approximately
the same amount from experiment and theory (i.e. essentially from the hadronic
contribution). Even after a projected reduction of the uncertainty by a factor
four both in $a_\mu({\rm exp})$ and $a_\mu({\rm SM})$ our numerical precision
is a factor ten below the uncertainty of the difference.

\section{Conclusions}

In this contribution we reported on the calculation of the four-loop QED
corrections to $a_\mu$ which involve closed electron
loops~\cite{Kurz:2016bau,Kurz:2015bia} [see $ A_2^{(8)}(m_\mu / m_e)$ in
Eq.~(\ref{eq::Amu})]. In Ref.~\cite{Kurz:2016bau} also the contribution
$A_3^{(8)}(m_\mu / m_e, m_\mu / m_\tau) $ with at least one electron and one
tau loop have been computed and the results for $A_2^{(8)}(m_\mu / m_\tau)$
can be found in Ref.~\cite{Kurz:2013exa}.  For all contributions perfect
agreement with the results of Ref.~\cite{Aoyama:2012wk} have been obtained.
The only missing four-loop contribution which still has to be cross-checked is
the universal part $A_1^{(8)}$.


\section*{Acknowledgments}

P.M. was supported in part by the EU Network HIGGSTOOLS PITN-GA-2012-316704.
The work of V.S. was supported by the Alexander von Humboldt Foundation
(Humboldt Forschungspreis).

\end{document}